


\font\bigbold=cmbx12
\font\ninerm=cmr9
\font\eightrm=cmr8
\font\sixrm=cmr6
\font\fiverm=cmr5
\font\ninebf=cmbx9
\font\eightbf=cmbx8
\font\sixbf=cmbx6
\font\fivebf=cmbx5
\font\ninei=cmmi9  \skewchar\ninei='177
\font\eighti=cmmi8  \skewchar\eighti='177
\font\sixi=cmmi6    \skewchar\sixi='177
\font\fivei=cmmi5
\font\ninesy=cmsy9 \skewchar\ninesy='60
\font\eightsy=cmsy8 \skewchar\eightsy='60
\font\sixsy=cmsy6   \skewchar\sixsy='60
\font\fivesy=cmsy5
\font\nineit=cmti9
\font\eightit=cmti8
\font\ninesl=cmsl9
\font\eightsl=cmsl8
\font\ninett=cmtt9
\font\eighttt=cmtt8
\font\tenfrak=eufm10
\font\ninefrak=eufm9
\font\eightfrak=eufm8
\font\sevenfrak=eufm7
\font\fivefrak=eufm5
\font\tenbb=msbm10
\font\ninebb=msbm9
\font\eightbb=msbm8
\font\sevenbb=msbm7
\font\fivebb=msbm5
\font\tensmc=cmcsc10


\newfam\bbfam
\textfont\bbfam=\tenbb
\scriptfont\bbfam=\sevenbb
\scriptscriptfont\bbfam=\fivebb
\def\Bbb{\fam\bbfam}

\newfam\frakfam
\textfont\frakfam=\tenfrak
\scriptfont\frakfam=\sevenfrak
\scriptscriptfont\frakfam=\fivefrak
\def\frak{\fam\frakfam}

\def\smc{\tensmc}


\def\eightpoint{%
\textfont0=\eightrm   \scriptfont0=\sixrm
\scriptscriptfont0=\fiverm  \def\rm{\fam0\eightrm}%
\textfont1=\eighti   \scriptfont1=\sixi
\scriptscriptfont1=\fivei  \def\oldstyle{\fam1\eighti}%
\textfont2=\eightsy   \scriptfont2=\sixsy
\scriptscriptfont2=\fivesy
\textfont\itfam=\eightit  \def\it{\fam\itfam\eightit}%
\textfont\slfam=\eightsl  \def\sl{\fam\slfam\eightsl}%
\textfont\ttfam=\eighttt  \def\tt{\fam\ttfam\eighttt}%
\textfont\frakfam=\eightfrak \def\frak{\fam\frakfam\eightfrak}%
\textfont\bbfam=\eightbb  \def\Bbb{\fam\bbfam\eightbb}%
\textfont\bffam=\eightbf   \scriptfont\bffam=\sixbf
\scriptscriptfont\bffam=\fivebf  \def\bf{\fam\bffam\eightbf}%
\abovedisplayskip=9pt plus 2pt minus 6pt
\belowdisplayskip=\abovedisplayskip
\abovedisplayshortskip=0pt plus 2pt
\belowdisplayshortskip=5pt plus2pt minus 3pt
\smallskipamount=2pt plus 1pt minus 1pt
\medskipamount=4pt plus 2pt minus 2pt
\bigskipamount=9pt plus4pt minus 4pt
\setbox\strutbox=\hbox{\vrule height 7pt depth 2pt width 0pt}%
\normalbaselineskip=9pt \normalbaselines
\rm}


\def\ninepoint{%
\textfont0=\ninerm   \scriptfont0=\sixrm
\scriptscriptfont0=\fiverm  \def\rm{\fam0\ninerm}%
\textfont1=\ninei   \scriptfont1=\sixi
\scriptscriptfont1=\fivei  \def\oldstyle{\fam1\ninei}%
\textfont2=\ninesy   \scriptfont2=\sixsy
\scriptscriptfont2=\fivesy
\textfont\itfam=\nineit  \def\it{\fam\itfam\nineit}%
\textfont\slfam=\ninesl  \def\sl{\fam\slfam\ninesl}%
\textfont\ttfam=\ninett  \def\tt{\fam\ttfam\ninett}%
\textfont\frakfam=\ninefrak \def\frak{\fam\frakfam\ninefrak}%
\textfont\bbfam=\ninebb  \def\Bbb{\fam\bbfam\ninebb}%
\textfont\bffam=\ninebf   \scriptfont\bffam=\sixbf
\scriptscriptfont\bffam=\fivebf  \def\bf{\fam\bffam\ninebf}%
\abovedisplayskip=10pt plus 2pt minus 6pt
\belowdisplayskip=\abovedisplayskip
\abovedisplayshortskip=0pt plus 2pt
\belowdisplayshortskip=5pt plus2pt minus 3pt
\smallskipamount=2pt plus 1pt minus 1pt
\medskipamount=4pt plus 2pt minus 2pt
\bigskipamount=10pt plus4pt minus 4pt
\setbox\strutbox=\hbox{\vrule height 7pt depth 2pt width 0pt}%
\normalbaselineskip=10pt \normalbaselines
\rm}


\def\pagewidth#1{\hsize= #1}
\def\pageheight#1{\vsize= #1}
\def\hcorrection#1{\advance\hoffset by #1}
\def\vcorrection#1{\advance\voffset by #1}

\newif\iftitlepage   \titlepagetrue               
\newtoks\titlepagefoot     \titlepagefoot={\hfil} 
\newtoks\otherpagesfoot    \otherpagesfoot={\hfil\tenrm\folio\hfil}
\footline={\iftitlepage\the\titlepagefoot\global\titlepagefalse
           \else\the\otherpagesfoot\fi}

\font\extra=cmss10 scaled \magstep0
\setbox1 = \hbox{{{\extra R}}}
\setbox2 = \hbox{{{\extra I}}}
\setbox3 = \hbox{{{\extra C}}}
\setbox4 = \hbox{{{\extra Z}}}
\setbox5 = \hbox{{{\extra N}}}



\def\ZZZ{{{\extra Z}}\hskip-\wd4\hskip 2.5 true pt{{\extra Z}}}
\def\Zed{\hbox{{\extra\ZZZ}}}       




\def\Z{{\Zed}}

\def\tr{\hbox{tr}}

\def\frac#1#2{{#1\over#2}}

\def\({\left(}
\def\){\right)}
\def\<{\langle}
\def\>{\rangle}

\def\pmb#1{\setbox0=\hbox{$#1$}%
   \kern-.025em\copy0\kern-\wd0
   \kern.05em\copy0\kern-\wd0
   \kern-.025em\raise.0433em\box0 }


\def\abstract#1{{\parindent=30pt\narrower\noindent\ninepoint\openup
2pt #1\par}}


\newcount\notenumber\notenumber=1
\def\note#1
{\unskip\footnote{$^{\the\notenumber}$}
{\eightpoint\openup 1pt #1}
\global\advance\notenumber by 1}


\global\newcount\secno \global\secno=0
\global\newcount\meqno \global\meqno=1
\global\newcount\appno \global\appno=0
\newwrite\eqmac
\def\romappno{\ifcase\appno\or A\or B\or C\or D\or E\or F\or G\or H
\or I\or J\or K\or L\or M\or N\or O\or P\or Q\or R\or S\or T\or U\or
V\or W\or X\or Y\or Z\fi}
\def\eqn#1{
        \ifnum\secno>0
            \eqno(\the\secno.\the\meqno)\xdef#1{\the\secno.\the\meqno}
          \else\ifnum\appno>0
            \eqno({\rm\romappno}.\the\meqno)
            \xdef#1{{\rm\romappno}.\the\meqno}
          \else
            \eqno(\the\meqno)\xdef#1{\the\meqno}
          \fi
        \fi
\global\advance\meqno by1 }


\global\newcount\refno
\global\refno=1 \newwrite\reffile
\newwrite\refmac
\newlinechar=`\^^J
\def\ref#1#2{\the\refno\nref#1{#2}}
\def\nref#1#2{\xdef#1{\the\refno}
\ifnum\refno=1\immediate\openout\reffile=refs.tmp\fi
\immediate\write\reffile{
     \noexpand\item{[\noexpand#1]\ }#2\noexpand\nobreak.}
     \immediate\write\refmac{\def\noexpand#1{\the\refno}}
   \global\advance\refno by1}
\def\semi{;\hfil\noexpand\break ^^J}
\def\nl{\hfil\noexpand\break ^^J}
\def\refn#1#2{\nref#1{#2}}

\def\ann#1#2#3{{\it Ann. Phys.} {\bf {#1}} (19{#2}) #3}

\def\jpA#1#2#3{{\it J.  Phys.} {\bf A{#1}} ({#2}) #3}

\def\np#1#2#3{{\it Nucl.  Phys.} {\bf B{#1}} (19{#2}) #3}
\def\pl#1#2#3{{\it Phys.  Lett.} {\bf {#1}B} (19{#2}) #3}

\def\prD#1#2#3{{\it Phys.  Rev.} {\bf D{#1}} (19{#2}) #3}
\def\prl#1#2#3{{\it Phys.  Rev.  Lett.} {\bf #1} (19{#2}) #3}


{

\refn\FaddeevA
{L. Faddeev and A.J. Niemi, 
\prl{82}{99}{1624}}

\refn\FaddeevB
{L. Faddeev and A.J. Niemi,
{\it Nature} 387 (1997) 58}

\refn\Gladikowski
{J. Gladikowski and M. Hellmund, \prD{56}{97}{5194};
R. Battye and P. Sutcliffe, hep-th/9811077, \prl{81}{98}{4798};
J. Hietarinta and P. Salo, \pl{451}{99}{60}}

\refn\tHooft
{G. 't Hooft,
\np{153}{79}{141};
{\it ibid.} {\bf B190} (1981) 455;
A. Polyakov, \np{120}{77}{429}}

\refn\Konishi
{K. Konishi and K. Takenaga, IFUP-TH-61-99, hep-th/9911097}

\refn\JackiwChrist
{N. Christ and R. Jackiw
\pl{91}{80}{228}}

\refn\Reinhardt
{H. Reinhardt,
\np{503}{97}{505};
C. Ford, U.G. Mitreuter, T. Tok and A. Wipf,
\ann{269}{98}{26}}

\refn\Jahn
{O. Jahn, \jpA{33}{2000}{2997}}

\refn\Witten
{E. Witten, 
\prl{38}{77}{121}}

\refn\FaddeevC
{L. Faddeev and A.J. Niemi, 
{\it From Yang-Mills Field to Solitons and Back Again}
(hep-th/9901037)}



}

\def\ve{\vfill\eject}

\def\veca{{\bf a}}
\def\vecb{{\bf b}}
\def\vece{{\bf e}}
\def\vecf{{\bf f}}
\def\vecx{\vec{\rm x}}
\def\hatx{\hat{\rm x}}

\def\vecn{{\bf n}}

\def\veca{{\bf a}}
\def\vecA{{\bf A}}
\def\d{{\rm d}}

\def\nnabla{\vec{\bigtriangledown}}
\def\nvece{\vec{\rm e}}




\pageheight{23cm}
\pagewidth{14.8cm}
\hcorrection{0mm}
\magnification= \magstep1
\def\bsk{%
\baselineskip= 15pt plus 1pt minus 1pt}
\parskip=5pt plus 1pt minus 1pt
\tolerance 6000


\null

{
\leftskip=100mm
\hfill\break
KEK Preprint 2000-18
\hfill\break
hep-th/0005064
\hfill\break
\par}

\smallskip
\vfill
{\baselineskip=18pt

\centerline{\bigbold 
Instantons, Monopoles
and the Flux Quantization}
\centerline{\bigbold 
in the Faddeev-Niemi Decomposition}

\vskip 30pt

\centerline{
\smc
Toyohiro Tsurumaru\note
{E-mail:\quad toyohiro.tsurumaru@kek.jp}
\quad {\rm and} \quad
Izumi Tsutsui\note
{E-mail:\quad izumi.tsutsui@kek.jp}
}

\vskip 5pt
{
\baselineskip=13pt
\centerline{\it
Institute of Particle and Nuclear Studies}
\centerline{\it
High Energy Accelerator Research Organization (KEK)
}
\centerline{\it Tsukuba, Ibaraki 305-0801, Japan}
}

\vskip 10pt

\centerline{
\smc
Akira Fujii\note
{E-mail:\quad fujii@het.phys.sci.osaka-u.ac.jp}
}

\vskip 5pt
{
\baselineskip=13pt
\centerline{\it
Department of Physics, Graduate School of Science}
\centerline{\it
Osaka University}
\centerline{\it Toyonaka, Osaka 560-0043, Japan}
}

\vskip 50pt

\abstract{%
{\bf Abstract.}\quad
We study how instantons arise in the low energy
effective theory of the $SU(2)$ Yang-Mills theory
in the context of the non-linear sigma model recently
proposed by Faddeev and Niemi.  We find a simple
relation between the instanton number $\nu$ and the charge 
$m$ of the monopole that appears in the effective theory.
It is given by $\nu = m \Phi/(2\pi)$, where $\Phi$
is the quantized flux associated with a $U(1)$
gauge field passing through the loop formed by
the singularity of the monopole.}

\bigskip
{\ninepoint

}
}

\hfill\break


\pageheight{23cm}
\pagewidth{15.7cm}
\hcorrection{-1mm}
\magnification= \magstep1
\def\bsk{%
\baselineskip= 14.7pt plus 1pt minus 1pt}
\parskip=5pt plus 1pt minus 1pt
\tolerance 8000
\bsk

\ve

\secno=1 \meqno=1


\noindent{\bf 1. Introduction}
\medskip

Yang-Mills (YM) theory is known to provide a basis
for describing fundamental interactions, most notably, 
the strong interaction.  In the high energy regime the 
theory admits perturbative studies thanks to 
the asymptotic freedom, whereas in the low energy regime
it defies a similar systematic analysis due to the strong
coupling behavior.  
Recently, Faddeev and Niemi made a
suggestion [\FaddeevA] that the $SU(2)$ YM theory 
$
{\cal L}_{{\rm YM}}=\frac1{2g^2}\tr F_{\mu\nu}F^{\mu\nu}
$
in a certain phase of the low energy
regime may be described by a modified version of the $O(3)$
nonlinear sigma model (NLSM) in four dimensions
often referred to as the Skyrme-Faddeev model, 
$$
{\cal L}_{{\rm SF}}
=\frac1{2\lambda^2}(\partial_\mu\vecn)^2
+\frac1{4e^2}(\epsilon^{abc}n^a\partial_\mu n^b\partial_\nu n^c)^2.
\eqn\NLSM
$$
This model permits stable soliton
solutions
[\FaddeevB,
\Gladikowski], which have been 
conjectured to be a candidate for glueballs~[\FaddeevA].

The key ingredient for establishing the 
connection between the $SU(2)$ YM theory
and the Skyrme-Faddeev
model is the so-called Faddeev-Niemi (FN)
decomposition\note{%
Notation: We denote $su(2)$-valued fields such as
$A = A^aT^a = A^a_\mu \d x^\mu T^a$ often by 
iso-vectors as 
$\vecA = (A^1, A^2, A^3)$.  Here $\d$ is the space-time
exterior derivative, and $T^a := \tau^a/(2i)$, 
where $\tau^a$
are the Pauli matrices,
is the $su(2)$ basis we use.
} 
$$
\vecA=C\,\vecn+(1+\sigma)\,\d\vecn\times\vecn
+\rho\,\d\vecn.
\eqn\FND
$$
This is a new version of the Abelian projection 
which decomposes the
YM field $\vecA$ 
into a vector field
$C$, a complex scalar 
$\phi = \rho+i\sigma$ and an
iso-vector field $\vecn$ of unit length $\sum_{a=1}^3
n^an^a=1$.   The vector field $C$ is a gauge field associated
with the $U(1)$ gauge transformations with respect to
$\vecn$, that is, it transforms as 
$C \rightarrow C + \d\theta$ 
under the $U(1)$ subgroup 
$U(x) = \exp[\theta\, n^aT^a]$ of $SU(2)$ transformations
for $\vecA$.
Thus, at on-shell, $C$ 
has two degrees of freedom when
gauge-fixed, and together with $\vecn,\ \rho$ and $\sigma$ 
it comprises the  
six degrees of freedom of the 
on-shell $SU(2)$ YM
field.
The original motivation for the Abelian projection 
proposed by 't Hooft and Polyakov~[\tHooft]  was to provide
a qualitative explanation for the color confinement by asserting
that in the low energy regime the YM field exhibits the dual
Meissner effect in a condensate of magnetic monopoles.
With regard to this, it is very
important to study how instantons --- a hallmark of
the non-perturbative aspect of the YM theory --- can affect
the magnetic monopole configurations 
in the Skyrme-Faddeev model obtained under 
the FN decomposition (\FND).
(The implication of the FN decomposition for confinement
and chiral symmetry breaking has been discussed in Ref.[\Konishi].)

In this note we point out that, when instantons are present
in the $SU(2)$ 
YM theory, monopoles must necessarily appear in
the low energy regime, {\it i.e.}, in the Skyrme-Faddeev model.
The relation between instantons and monopoles has 
been discussed earlier by Christ and Jackiw for
physically static field configurations [\JackiwChrist], 
where the instanton number was shown to coincide with 
the monopole charge for $SU(2)$.  In the context of 
the original Abelian projection, 
a similar relation has been found in [\Reinhardt] for generic
configurations under the Weyl gauge.
More recently, it has been shown without
reference to gauge fixing that 
the instanton number is
given, modulo the monopole charge, by
the  (generalized) Hopf invariant associated with
an adjoint scalar field [\Jahn]. 
Here we examine the problem in the context of 
the FN decomposition, and argue that
the relation between the instanton number and
the monopole charge is again given by a similar
formula, involving this time a quantized 
flux trapped by the
monopole. Specifically, 
if the original YM gauge field that admits
the FN decomposition possesses a nonzero instanton
number, the corresponding vector field $\vecn$ 
must have singularity in closed
circles, or `monopole loops'.  We shall find that
the flux 
$\Phi$ associated with the
$U(1)$ gauge field $C$ trapped by the monopole loop must be
quantized topologically, and that the instanton number $\nu$ is
given by the product of the monopole charge $m$ and 
the topological integer of the flux 
$\Phi$ (see Fig.1),
$$
\nu = m\, {\Phi\over{2\pi}}.
\eqn\nuMW
$$
The rest of the paper is devoted to show this simple
relation.

\topinsert
\let\picnaturalsize=N
\def\picsize{5.5cm}
\def\picfilename{FN_Fig1.eps}
\ifx\nopictures Y\else{\ifx\epsfloaded Y\else\input epsf \fi
\global\let\epsfloaded=Y
\ifx\picnaturalsize N\epsfxsize \picsize\fi
\hskip 2.5cm\epsfbox{\picfilename}}\fi
\vskip 0cm
\abstract{%
{\bf Figure 1.} 
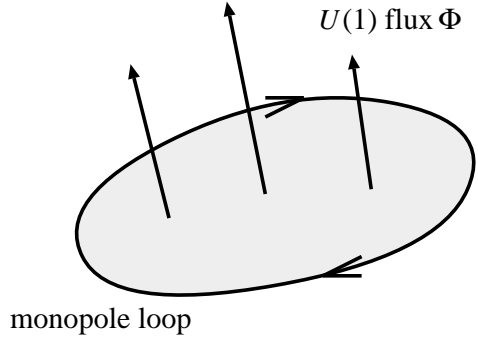
The singularity of $\vecn$ arises as a closed loop (monopole loop) 
in the
spacetime for non-vanishing instanton sectors.
The instanton number $\nu$ 
is given by the monopole charge $m$ times the 
integer of the quantized flux $\Phi$
associated with 
the $U(1)$ gauge field $C$ passing through the surface 
encircled by the
monopole loop.
}
\endinsert


\secno=2 \meqno=1


\bigskip
\noindent{\bf 2. Instanton and singularity}
\medskip

Prior to the discussion on the relation (\nuMW), we first
show that, under the FN decomposition (\FND), the field
$\vecn$ is necessarily singular if the original $\vecA$
possesses a non-vanishing instanton number.  
Following the conventional procedure, we assign 
a configuration of the gauge field on our
spacetime diffeomorphic to $S^4$ by introducing
a set of local coordinate patches $V_k \simeq D^4$ for 
$k = 1$, $2$, say.  Let  $A_k$ be 
the gauge potential on the patch 
$V_k$.   On the overlap
$V_1\cap V_2$, the gauge fields are related by
$$
A_2 = U^\dagger A_1 U + U^\dagger \d U\ ,
\eqn\gtrarel
$$
with the transition function $U(x) \in SU(2)$.
The 
instanton number $\nu$ of the gauge field is 
encoded in the transition function
in its winding
number on the boundary 
$\partial V_1\simeq S^3$
(or equivalently on $\partial V_2$), 
$$
\nu= -{1\over{8\pi^2}}\int_{S^4}  \tr(F\wedge F)
=\frac1{24\pi^2}
\int_{\partial V_1}
\tr(U^\dagger \d U)^3.
\eqn\instanton
$$

If we now adopt the FN decomposition to each of
the gauge fields on the patches, we have 
$$
\vecA_{k}=C_{k}\vecn_k
+(1+\sigma_k)\,\d\vecn_k\times\vecn_k
+\rho_k\,\d\vecn_k\ ,
\eqn\gaugetwo
$$
for $k = 1$, 2.  We remark 
that the FN decomposition implies, in effect, 
a specific gauge fixing and use of 
an on-shell condition for
$\vecA$  (except for the part of the $U(1)$ gauge field $C$), 
and therefore it is a nontrivial question if the
decomposition (\gaugetwo) is actually available over
the entire
patches we have introduced.  In what follows we consider the
class of configurations for which this is possible, 
knowing that standard instanton configurations such as
Witten's solution (see below) belong to this class.
Further, for our effective picture of
the Skyrme-Faddeev model which has no
local gauge symmetry, we need to
regard the field $\vecn$ {\it physical}, 
that is, it represents
the low energy dynamical freedom of the YM theory 
irrespective of the gauge chosen.  In other words, 
we assume 
that  {\it given a configuration of the 
$SU(2)$ gauge field $\vecA$ with gauge
equivalent configurations identified, the corresponding
field $\vecn$ in the FN decomposition can be uniquely 
determined up to a global $O(3)$ transformation.}
This is a crucial but necessary demand for the FN
decomposition to work in practice, allowing us to 
discuss the topological aspect of the YM theory
in terms of the field $\vecn$. 

Upon this assumption we can perform, 
with no loss
of generality, a global $O(3)$ transformation so that
we have 
$\vecn_1=\vecn_2$ on the overlap $V_1\cap V_2$. 
Thus 
the transition function $U$ defined on the 
overlap must
be of the form,
$$
U({x})=\exp\left[\theta(x) n(x)\right],\qquad
n(x) := \sum_an^a(x)T^a\ ,
\eqn\DefTheta
$$
with some
function $\theta(x)$ on $V_1\cap V_2$.
Then it is easily seen that if $\vecn$ is
regular  everywhere on $\partial V_1$, 
the winding number (\instanton) of $U({x})$ must be zero.
Indeed, if we expand $U({x})$ in (\DefTheta) in the form,
$$
U({x})=\alpha(x) + \beta(x)\, n(x)\ ,
\eqn\Gx
$$
with
$$
\alpha(x) = \cos {\theta\over2} = 
{1\over 2}{\rm tr}\,U\ ,
\qquad
\beta(x) = \sin {\theta\over2} = -2\,{\rm tr}\,(U,\vecn)\ ,
\eqn\supGx
$$
then $\alpha(x)$ and $\beta(x)$ are 
also regular for regular $n(x)$, 
because $U(x)$ is given regularly over the overlap.
Since $(\alpha,\beta)$ may be regarded as a map from 
$\partial V_1 \simeq S^3$ to 
$S^1$, it can be deformed continuously to the constant map
$(\alpha,\beta)=(1,0)$ on account of $\pi_3(S^1)=0$.  
For regular $\vecn$, this means
that 
$U({x})$ can be continuously deformed to the constant map
$U({x})=1$, for which the winding number vanishes. 
It follows therefore that 
$\vecn$ must necessarily be singular if the gauge field
$\vecA$ has a non-vanishing instanton number.
Note that the regularity of
$U(x)$ demands 
that the parameter $\theta$ in
(\DefTheta) be such that the function
$\beta$ in (\Gx) converges to zero 
sufficiently fast
when $\vecn$ approaches its singularity. 
Thus, for singular $\vecn$, 
the functions $(\alpha, \beta)$
are still given regularly, even though the deformation
of $(\alpha,\beta)$ mentioned above
cannot be performed due to
the constraint $\beta = 0$ at the singularity.

The singularity of $\vecn$ can be classified by
the topological property $\pi_2(S^2)=\Z$ by
regarding $\vecn$ as a map from 
an arbitrary two-dimensional sphere $S^2$ in
the spacetime to the target $S^2$. 
The topological integer can be provided
by
$$
m
=\frac{1}{2\pi}\int_{S^2}\tr(n\, \d n\wedge\d n)\ ,
\eqn\winding
$$
which we call the \lq monopole charge'.
In the four-dimensional spacetime, 
we expect that the singularity of $\vecn$ forms 
a one-dimensional closed loop [\tHooft], {\it i.e.}, 
a monopole loop as shown in Fig.1.  From
the foregoing argument we find that, if 
$\nu \ne 0$
then $m \ne 0$, or equivalently, if $m = 0$ then $\nu = 0$.

\topinsert
\vskip 1.5cm
\let\picnaturalsize=N
\def\picsize{10cm}
\def\picfilename{FN_Fig2.eps}
\input epsf
\ifx\picnaturalsize N\epsfxsize \picsize\fi
\hskip 0cm\epsfbox{\picfilename}
\vskip 0.5cm
%
%
\abstract{%
{\bf Figure 2.}
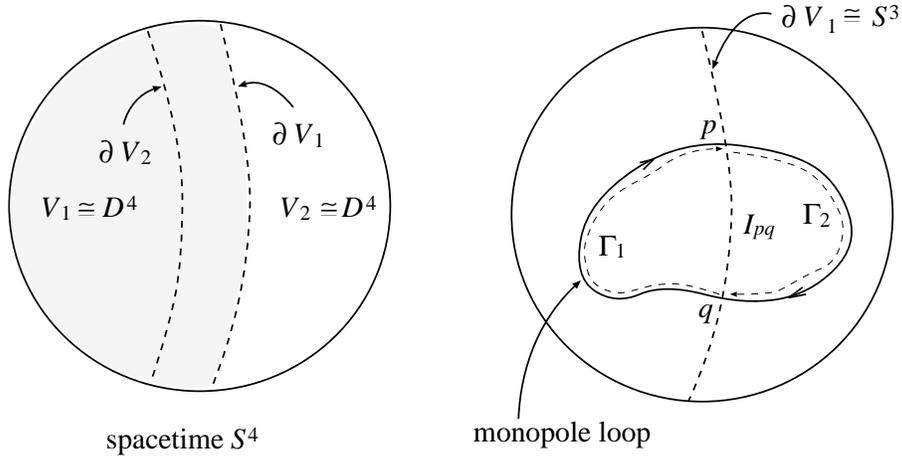
The spacetime $S^4$ is covered with two local patches 
$V_1$, $V_2\simeq D^4$.
The monopole loop lies across the overlap
$V_1\cap V_2$, and hence can be cut into two
pieces $\Gamma_1$ and $\Gamma_2$ at the two intersections
$p$ and $q$ with the boundary $\partial V_1$.
}
\endinsert

At this point, it is instructive for us to
look at explicit solutions of the YM instantons
that admit the FN decomposition.  For this we
take Witten's ansatz [\Witten] which
is an axially symmetric self-dual solution to
the $SU(2)$ YM equation with arbitrary instanton numbers.
With the coordinate $(\vecx,t)$ on a local patch, the ansatz
assumes for the solution the most general form which is invariant 
under $SO(3)$ rotations,
$$
A_0^a=\frac{A_0x^a}{r}\ ,
\qquad
A_j^a=\frac{\varphi_2+1}{r^2}\epsilon_{jak}x_k+
\frac{\varphi_1}{r^2}[\delta_{ja}r^2-x_jx_a]+
A_1\frac{x_jx_a}{r^2}\ , 
\eqn\anzats
$$
where $\varphi,A_0,A_1$ are functions of the radius $r=|\vecx|$
and $t$.
Note that
this is already in the form of the FN
decomposition (\FND), as seen by setting $\varphi_1=\rho,\
\varphi_2=\sigma,\ C_i=(x_i/r)A_1,\ C_0=A_0$ and
$n^a(\vecx)=x^a/r$ [\FaddeevC], where now the monopole loop
is located at the origin of the space.  If we let $z = r+it$
and define $\psi$ by
$A_\mu=\epsilon_{\mu\nu}\partial_\nu\psi$ ($\mu,\nu=0,1$ where 
0 and 1 refers to $t$ and $r$, respectively), self-dual solutions
with instanton number $\nu = k -1 $ are obtained by
$\psi=-\ln[({1-g^*g})/(2r)]$ with
$g(z)=\prod^{k}_{i=1}({a_i-z})/({a_i^*+z})$
using arbitrary complex constants $a_i$ satisfying 
${\rm Re}\,a_i>0$.
Clearly, the instanton number
$\nu$ is not determined (and hence can be put even to zero)
under the field
$n^a(\vecx,t)=x^a/r$ which is singular at the
origin and has the monopole 
charge $m=1$ (see
(\winding)).

The above example shows that the instanton number $\nu$
is not determined by the monopole charge alone,
and that to account for the instanton
number the configuration of the $U(1)$
gauge field $C$ must also be considered.  Indeed, for the
ansatz (\anzats) the instanton number (\instanton) reduces to
$$
\nu=\frac1{2\pi}\int_{D^2} \d^2x\left[\ \partial_\mu(\epsilon_{ij}
\epsilon_{\mu\nu}\varphi_iD_\nu\varphi_j)
+\frac12\epsilon_{\mu\nu}F_{\mu\nu}\right]\ ,
\eqn\InstantonTwo
$$
where $D^2$ is a two-dimensional disc whose boundary
is the monopole loop.  We find that 
the second term yields the flux of the
gauge field $C$ passing through the monopole loop, and
this will be seen the only 
contribution to the instanton number $\nu$ ({\it i.e.},
the first term given by the surface integral vanishes)
on a general basis shortly.

\topinsert
\let\picnaturalsize=N
\def\picsize{6cm}
\def\picfilename{FN_Fig3.eps}
\input epsf
\ifx\picnaturalsize N\epsfxsize \picsize\fi
\hskip 2.8cm\epsfbox{\picfilename}
\vskip 0.5cm
\abstract{%
{\bf Figure 3.}
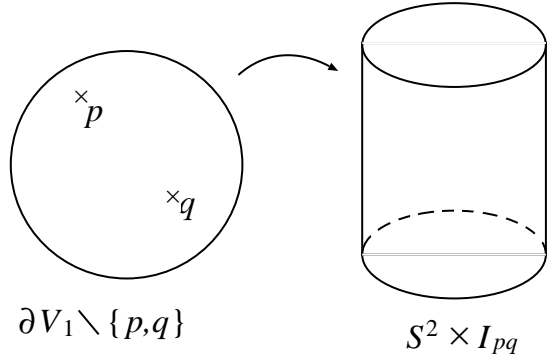
The removal of the two points $p$ and $q$ 
from $\partial V_1$ yields the cylinder
$\partial V_1\setminus\{p, q\} \simeq S^2\times I_{pq}$.
}
\endinsert

\secno=3 \meqno=1


\bigskip
\noindent{\bf 3. Instanton number vs monopole charge}
\medskip

The formula (\InstantonTwo) suggests 
that the flux associated with $C$ penetrating through
the surface 
encircled by 
the monopole loop contributes to the instanton
number
$\nu$.
We now show that $\nu$ is in fact proportional to
the flux in just the way (\nuMW) we mentioned in the
Introduction.  We do this in two steps, first
showing that the instanton number is proportional 
to the monopole charge, and then arguing that
the proportional factor is given by the flux which
is quantized.

To this end,
recall that for nonzero $\nu$ the field $\vecn$ is
singular on the boundary $\partial V_1$.  This implies that
the monopole loop (which is the line of singularity of
$\vecn$) must intersect with 
$\partial V_1$ 
twice or more generally even times.  Suppose, for simplicity,
that there are two intersections on $\partial V_1$,  
which we denote as $p$ and $q$. (When there are more
than two intersections we may choose a different
set of coordinate patches so that the boundary
$\partial V_1$ intersects with the monopole loop 
only twice.)
From (\supGx) we find that in the overlap
$\partial V_1 \cap \partial V_2$ the function 
$\theta$ can be defined
regularly.
At the singular points $p$ and $q$, 
we have $\beta = \sin(\theta/2)=0$ 
to ensure the regularity of
$U$ and hence
$$
\theta = {2\pi} \times \hbox{integer}\ ,
\eqn\alphaBC
$$
at both $p$ and $q$.  

Consider then the cylinder 
$\partial V_1\setminus\{p, q\} 
\simeq S^2\times I_{pq}$ (without edges)
obtained by removing the two points $p$ and
$q$ from 
$\partial V_1 \simeq S^3$, 
where $I_{pq}$ is an interval from
$p$ to $q$ on the cylinder (see Fig.3).
Since $U(x)$ is regular, the removal of the two points from
the domain of integration for the instanton
number (\instanton) does not alter the
outcome.  Thus, instead of the domain $\partial V_1 \simeq S^3$
we may use the cylinder to evaluate the instanton number
(\instanton) as 
$$
\nu 
=\frac1{24\pi^2}\int_{S^2\times I_{pq}}
\tr\,(U^\dagger\d U)^3 
=\frac{1}{4\pi^2}\int_{S^2\times I_{pq}}
\left(\,1-\cos\theta\,\right)
\d\theta\,\, \tr\,(n\d n\wedge\d n)\ ,
\eqn\instantonthree
$$
where we have used
$$
U^\dagger \d U=\left(n\,\d\theta+\sin\theta\,\d n\right)
-[n,\d n](1-\cos\theta)\ .
\eqn\no
$$
Let us choose the coordinates of the cylinder 
$S^2\times I_{pq}$ such that $\theta$ is
constant on $S^2$ at each point of $I_{pq}$. 
This choice of coordinates allows us to evaluate the integral
(\instantonthree) by 
two separate integrals over $\theta$ and $\vecn$ as 
$$
\nu=\frac{m}{2\pi}\int_{I_{pq}}
\left(\,1-\cos\theta\,\right)\d\theta
=\frac{m}{2\pi}\int_{I_{pq}}\d\theta\ ,
\eqn\instantonfour
$$
where $m$ is the monopole charge (\winding), and we have used
(\alphaBC) for the second equality.  The condition (\alphaBC)
also ensures that the r.h.s.~of 
(\instantonfour) gives an integer as required. 


\secno=4 \meqno=1


\bigskip
\noindent{\bf 4. Flux quantization}
\medskip

It remains to show that the integer factor multiplying
$m$ in (\instantonfour) is the flux $\Phi$ associated with
the $U(1)$ gauge  
field $C$ passing through the monopole loop.
Let $\Gamma_k$, $k = 1$, 2, 
be the contours obtained by
cutting the monopole loop in half at
$p$ and $q$ (see Fig.2).  Then the
flux $\Phi_k$ penetrating the surface encircled
by $\Gamma_k$ and $I_{pq}$ is given by
$\Phi_k = (\int_{\Gamma_k} + \int_{I_{pq}}) C_k$.
Noting that $C_1$ and $C_2$ are related by the relation 
$C_2
= C_1+ \d \theta$ on $I_{pq}$, 
we find that the total flux is given by
$$
\Phi= \Phi_1 + \Phi_2 = 
\int_{\Gamma_1}C_1 + 
\int_{\Gamma_2}C_2
+\int_{I_{pq}}\d\theta\ .
\eqn\flux
$$
We shall argue that the component of the $U(1)$ field 
$C_k$ along the monopole loop 
actually vanishes at the monopole loop, and hence 
the contributions from the two 
contours $\Gamma_1$ and $\Gamma_2$ in (\flux) disappear.

For this, we choose a local coordinate patch in 
$V_1$ and parameterize it by $(\vecx,t)$, where the
origin $\vecx = \vec0$ is taken to be at the monopole loop, 
and $t$ is the local coordinate along the monopole loop.
With $\hatx = \vecx/\vert \vecx \vert$ and 
$r = \vert \vecx \vert$, 
we consider the limit,
$$
\vecn_0(\hatx,t):=\lim_{r\to 0}\vecn(r\hatx,t)\ .
\eqn\limit
$$
Here we suppose that 
$\vecn(x)$ has a certain limit dependent on
the direction $\hatx$.  To specify the direction,
let us use the polar coordinates 
$(r, \vartheta, \varphi)$ and
consider the unit vectors $\nvece_r$,
$\nvece_\vartheta$,
$\nvece_\varphi$ associated with them. 
(Note that
$\vecn$ has a direction-independent limit 
if and only if it
is regular, for which we have 
$m = 0$ and hence the relation
(\nuMW) holds trivially.)
For $\vecn$ with $m \ne 0$, the radius of
the sphere $S^2$ on which the monopole charge (\winding) 
is evaluated can be taken as small as we
wish without changing the value $m$.   From this
we recognize that the derivative,
$$
\nnabla\vecn=\nvece_r\partial_r\vecn
+\frac{\nvece_\vartheta}{r}\partial_\vartheta\vecn
+\frac{\nvece_\varphi}{r}\partial_\varphi\vecn\ ,
\eqn\derv
$$
diverges as $1/r$ on average, or
more precisely, on an area of finite
volume on the $S^2$. Under the existence of the limit (\limit),
we obtain
$$
\lim_{r\to
0}\frac{|\partial_t\vecn|}{\left|\nnabla\vecn\right|}=
\lim_{r\to 0}\frac{|\partial_t\vecn_0|}
{\left|\nnabla\vecn\right|}
=0\ ,
\eqn\limittwo
$$
where we have used the notation 
$\left|\nnabla\vecn\right| 
= \sqrt{\sum_{i, a} (\partial_i n^a)^2}$. 
On the other hand, from
the inequality,
$$
|(1+\sigma)\partial_t\vecn\times\vecn+\rho\partial_t\vecn|
=\left|(1+\sigma)\nnabla\vecn\times\vecn+\rho\nnabla\vecn\right|
\frac{|\partial_t\vecn|}{\left|\nnabla\vecn\right|}
\le|\vec{\bf A}|
\frac{|\partial_t\vecn|}{\left|\nnabla\vecn\right|}\ ,
\eqn\partialn
$$
and the fact that $|\vec{\bf A}|$ is finite,
we observe that the l.h.s.~of (\partialn) 
converges to zero.   
Hence it follows
that
$$
\vecA_t\to C_t\vecn\qquad{\rm for}\quad r\to 0\ .
\eqn\limA
$$
However, since $\vecA_t$ is smooth while $\vecn$ is singular at
the origin $r = 0$, we must have
$\lim_{r \to 0}\vecA_t=0$ and
$$
C_t\to0\qquad{\rm for}\quad r\to 0\ .
\eqn\limitC
$$

Thus we see that the component $C_t$ along the
monopole loop vanishes identically at the monopole
loop.  This in turn implies that the flux $\Phi$ in (\flux)
is quantized as $\Phi = 2\pi \times \hbox{integer}$,
where the integer is given
by the difference in the integers (\alphaBC) 
of the angle $\theta$ at the singular points
$p$ and $q$.  (The quantization condition
takes a more familiar form $\Phi = 2\pi\hbar/g \times
\hbox{integer}$
if the flux is evaluated for $C/g$ with a properly 
rescaled $g$.)
In conclusion, we have shown that the relation
(\nuMW) holds for the class of those 
$\vecn$ for which the
limit (\limit) exists.
We note that Witten's
ansatz for instantons has such a limit, and we expect that 
any physically interesting configurations will also have it, because 
violent fluctuations will be smeared out in
the low energy regime.  For completeness, however, in the Appendix
we shall provide an outline of the argument for more
generic configurations for $\vecn$.
\bigskip

\noindent
{\bf Acknowledgement:}  
The authors wish to thank A.J.~Niemi for 
helpful discussions.  This work is supported in part 
by the Grant-in-Aid for Scientific Research (C)
under Contract No.~11640301 provided by
the Ministry of Education, Science, Sports and
Culture of Japan.

\ve
\secno=0 \appno=1 \meqno=1 


\vskip 1cm 

\noindent{\bf Appendix} 

We here discuss the general case in which $\vecn$
may not have the limit (\limit).
Let us regard $A_\mu^a$ as a four times three matrix and
consider the quantity 
${\rm rank}\,\vecA$ which is the number of linearly
independent (column or, equivalently, row) 
vectors in the matrix $A_\mu^a$.
Choose a point on the monopole loop to which we assign 
the value $t = 0$.  Consider then a neighbourhood
of the point with four-dimensional radius
$\epsilon$ such that   
$$
{\rm rank}\vecA(x)\ge {\rm rank} \vecA(x=0)\qquad
{\rm for}\quad |x|=\sqrt{x_\mu x^\mu}<\epsilon.
\eqn\apone
$$
Since 
${\rm rank}\,\vecA(x=0)=0$ means 
$\vecA(x=0)=0$ and hence $\lim_{r\to0}C_\mu=0$, 
we only need to consider the cases where 
${\rm rank}\,\vecA\ge1$ at
$x=0$.

\medskip
\noindent{{\bf (i)} $\hbox{rank}\,{\vecA}(x=0) = 3$:}
In this case we have 
$k^\mu(x)$
satisfying
$k^\mu\vecA_\mu=0$ given by
$$
k^\mu := \frac1{3!}
\epsilon^{\mu\alpha\beta\gamma}\epsilon_{abc}
A^a_\alpha A^b_\beta A^c_\gamma\ ,
\eqn\no
$$
which is a non-vanishing smooth function 
for ${\rm rank}\,\vecA=3$. 
As can be easily seen from 
(\FND), we have 
$(1+\sigma)^2+\rho^2\ne0$ for $r\ne0$, and hence 
$k^\mu\vecA_\mu=0$ means that
$$
k^\mu\partial_\mu\vecn=0\ .
\eqn\wno
$$
Thus $k^\mu$ is a Killing vector for $\vecn$, and hence
it must point to the $t$-direction along the monopole
loop at $x = 0$.  It then follows that 
$\vecA_t(x=0) = 0$ and, therefore, $C_t(x=0) = 0$.

\medskip
\noindent{{\bf (ii)} $\hbox{rank}\,{\vecA}(x=0) = 2$ and 1:}
For $\vecA_\mu$ with rank $= 2$,
we can put $\vecA_\mu$ and $\partial_\mu\vecn$ 
(which also has rank $= 2$ because $\vecn$ has two
independent freedoms) in the form,
$$
\vecA_\mu(x) 
=\alpha_\mu\veca+\beta_\mu\vecb+O(|x|)\ , \qquad
\partial_\mu\vecn(x)
= \xi_\mu(x)\vece(x) +\chi_\mu(x)\vecf(x)\ ,
\eqn\eone
$$
where $\alpha_\mu$,
$\beta_\mu$, $\veca$, $\vecb$ are constant
vectors while other vectors $\xi_\mu$,
$\chi_\mu$, $\vece$, $\vecf$
are coordinate
dependent.   Plugging these into the identity,
$$
(\vecA_\mu\times\vecn)\times\vecn
= - (1+\sigma)\,\partial_\mu\vecn\times\vecn
+\rho\,\partial_\mu\vecn \ ,
\eqn\no
$$
which holds for
$r>0$, we see that
the vectors $\xi_\mu(x)$ and 
$\chi_\mu(x)$ are given by a linear combination
of $\alpha_\mu$ and $\beta_\mu$ with coordinate
dependent coefficients.  We thus find that, 
for $|x|\to0$, $\partial_\mu\vecn$ takes the form, 
$$
\partial_\mu\vecn\to\alpha_\mu{\bf u}
(\vecn)+\beta_\mu{\bf v}(\vecn),
\eqn\no
$$
where ${\bf u}(\vecn)$, ${\bf v}(\vecn)$ are
vectors orthogonal to $\vecn$.  Let 
$\gamma_\mu$ be a vector which is orthogonal to both
$\alpha_\mu$ and $\beta_\mu$ and has no time-component.
If we choose the $z$-direction of space to be parallel
to the vector $\gamma_\mu$, then we get
$$
\lim_{|x|\to
0}\frac{|\partial_z\vecn|}{\left|\nnabla\vecn\right|}
=0\ .
\eqn\limittwo
$$
This is, however, impossible 
because $\partial_i\vecn$ must diverge uniformly
for all $i$ as $r \to 0$ if $\vecn$ has
a non-vanishing monopole charge $m \ne 0$.
Thus, we conclude 
that $\vecA_\mu$ with rank $= 2$ cannot arise at 
the monopole loop.  The case $\vecA_\mu$ with rank $= 1$
can also be denied by a similar argument.

\ve
\baselineskip= 15.5pt plus 1pt minus 1pt
\parskip=5pt plus 1pt minus 1pt
\tolerance 8000
\vfill\eject

  \vfill\eject\immediate\closeout\reffile
  \centerline{{\bf References}}\bigskip\frenchspacing%
  \input refs.tmp\vfill\eject\nonfrenchspacing
\bye